\documentclass[aps,
prl,
reprint,
superscriptaddress,
longbibliography]{revtex4-2}
\usepackage{mathrsfs}
\usepackage{amsmath}
\usepackage{amssymb}
\usepackage{bm}
\usepackage{booktabs}
\usepackage{siunitx}
\usepackage{subfig}
\usepackage{graphicx}
\usepackage[colorlinks=true,linkcolor=blue,citecolor=blue,urlcolor=blue]{hyperref}

\graphicspath{{figures/}}

\newcommand{\dd}{\mathrm d}
\newcommand{\ii}{\mathrm i}
\newcommand{\ee}{\mathrm e}

\newcommand{\ket}[1]{\left|#1\right\rangle}
\newcommand{\bra}[1]{\left\langle #1\right|}

\newcommand{\kvec}{\mathbf k}
\newcommand{\pvec}{\mathbf p}
\newcommand{\Pvec}{\mathbf P}
\newcommand{\Rvec}{\mathbf R}
\newcommand{\rvec}{\mathbf r}

\newcommand{\Bvec}{\mathbf B}
\newcommand{\AMp}{OAM partition}
\newcommand{\rOAM}{relative OAM}
\newcommand{\cmOAM}{CM's OAM}
\newcommand{\eOAM}{electron's OAM}
\usepackage{todonotes}

\begin{document}

\title{Orbital-angular-momentum partition in hydrogen photoionization by a monochromatic vortex beam}

\author{Zhongchen Xing}
\email{xingzhongchen@pku.edu.cn}
\affiliation{State Key Laboratory for Mesoscopic Physics and Collaborative Innovation Center of Quantum Matter, School of Physics, Peking University, Beijing 100871, China}

\author{Chengyin Wu}
\affiliation{State Key Laboratory for Mesoscopic Physics and Collaborative Innovation Center of Quantum Matter, School of Physics, Peking University, Beijing 100871, China}

\author{Zheng Li}
\email{zhengli@wit.edu.cn}
\affiliation{School of Science and Hubei Key Laboratory of Optical Information and Pattern Recognition, Wuhan Institute of Technology, Wuhan 430205, China}
\affiliation{Wuhan National Laboratory for Optoelectronics and School of Physics, Huazhong University of Science and Technology, Wuhan 430074, China}
\author{Marcelo F. Ciappina}
\email{marcelo.ciappina@gtiit.edu.cn}
\affiliation{Department of Physics, Guangdong Technion - Israel Institute of Technology, Shantou, Guangdong, China}
\affiliation{Technion – Israel Institute of Technology, Haifa, Israel}
\affiliation{Guangdong Provincial Key Laboratory of Materials and Technologies for Energy Conversion, Guangdong Technion - Israel Institute of Technology, Shantou, Guangdong, China}
\date{\today}

\begin{abstract}

Understanding how optical orbital angular momentum (OAM) is transferred to matter requires treating recoil and translational motion alongside the internal electronic dynamics. We develop a center-of-mass-resolved theory of one-photon ionization of hydrogen by a monochromatic Laguerre--Gaussian beam and show that the Bessel-vortex photoelectron predicted in fixed-target models is a preparation-dependent limit. For a sharply defined atomic center-of-mass momentum, the recoil records the photon-cone azimuth, and tracing over it generally destroys the coherence required for a pure electron vortex. In the small-transverse-retardation regime, the optical OAM is transferred predominantly to the center-of-mass motion and hence, in the laboratory frame, to the proton. Finite-retardation corrections redistribute angular momentum between center-of-mass and relative motion, while an additional correlation contribution to the electron and proton angular momenta can be tuned through the spatial uncertainty of the atomic center of mass. These results reveal atomic recoil as a key element of OAM transfer in photoionization.

\end{abstract}

\maketitle

\newpage


Optical vortex beams carry orbital angular momentum (OAM) through their
azimuthally varying phase structure~\cite{Beijersbergen1994,Shen2019}. For a
paraxial Laguerre--Gaussian (LG) mode, this phase is commonly written as
$\exp(\ii\ell\phi)$, where the integer $\ell$ is the topological charge (TC) and an
ideal mode carries an OAM of $\ell\hbar$ per photon~\cite{Allen:92}. The family
of structured fields is broader than the standard LG basis and includes, among
others, Bessel--Gauss, Mathieu, elegant LG, and perfect optical-vortex
beams~\cite{Wang:23,Loxpez-Mariscal:06,Martinez-Castellanos:13,Vaity:15,das2026review}. Their
spatial amplitude, phase, and polarization provide additional degrees of
freedom with which to control light--matter interactions. This control has led
to applications ranging from optical manipulation and imaging to
OAM-multiplexed communication~\cite{padgett:11,Otomo:14,wang:12}, and has also
enabled the generation and control of vortex radiation in nonlinear and
strong-field processes~\cite{Kong:17,Bikash:24}.

The exchange of angular momentum between structured light and matter is a
long-standing and well-studied subject. Optical OAM and helicity can be transferred directly to macroscopic
objects, as demonstrated through the rotation and mechanical manipulation of
trapped particles~\cite{He1995,Friese1996,Porta2004}. At the microscopic level, however,
the statement that a photon carries $\ell\hbar$ of angular momentum does not by itself
specify which target degree of freedom receives it. The answer depends on
the light--matter approximation, on the spatial extent and preparation of the
target, and on which final degrees of freedom are observed or traced out.
Weak-field selection-rule analyses make this distinction explicit: at electric
dipole order, the optical OAM may enter primarily through the center-of-mass
(CM) coordinate, whereas higher multipoles or field-gradient terms are needed
for the transverse phase structure to couple directly to internal
motion~\cite{Babiker2002,Alexandrescu2006,Mondal2014,Maslov2024}. Experiments with trapped
ions have, correspondingly, demonstrated both transfer of optical OAM to a
bound electron and coherent transfer of transverse optical momentum to the
ion's motion~\cite{Schmiegelow2016,Stopp2022}. Free translational motion can itself support atomic vortex states, emphasizing that the CM is a genuine physical channel for OAM rather than a passive spectator in the dynamics~\cite{Andersen2006,Piovella2022,Gisbert2022}.

Photoionization provides a particularly transparent setting in which to study
this partition. In photoionization driven by plane-wave photons, the energy and angular
distribution of the emitted electron encode the structure and dynamics of the
target. With a vortex beam, the transverse intensity profile, the azimuthal
phase, and the position of the atom relative to the beam axis introduce
additional spatial information. Calculations for hydrogen-like ions have shown
that photoelectron angular distributions can depend strongly on the impact
parameter when the atom is close to the vortex core~\cite{Matula:13}, while
related selection-rule studies have clarified how twisted-light absorption
depends on target localization~\cite{scholz2014absorption}. Extensions to
molecular ions and macroscopic multielectron targets have demonstrated that
the molecular scale, target averaging, and the beam geometry can substantially
modify the observable photoelectron distribution~\cite{Peshkov:15,Kiselev:23}.
In the strong-field regime, the emitted electron can itself carry structured
OAM. In particular, strong-field ionization driven by
counter-rotating circularly polarized fields has been shown to generate vortex
structures in the photoelectron momentum distribution, which can be understood
as interference between twisted electron states with different orbital angular
momenta~\cite{Maxwell2021}. Twisted attosecond pulses have also been proposed
as probes of the spectral and energy-flow structure of vortex fields~\cite{Muller:16,Surzhykov:16},
and counter-rotating circularly polarized attosecond pulses provide another
route to photoelectron vortices~\cite{Zhang:24}.

Of particular relevance for the present work, fixed-target treatments of atomic
photoionization by ideal Bessel or LG photons predict the transfer of the optical
vortex phase to the outgoing electron when a single atom is located on the
beam axis~\cite{Pavlov:24,Das2025}. In that limit, the final electronic
wavefunction can have a definite OAM projection and a Bessel-like transverse
profile with a central node and concentric rings. The use of an ideal Bessel
beam is analytically convenient but physically idealized because such a field
has infinite transverse extent and energy. A high-radial-index LG beam offers
a realizable alternative: in the appropriate asymptotic regime its transverse
spectrum approaches a narrow Bessel ring and its spatial profile reproduces
the relevant Bessel-like structure over a finite region~\cite{Guo:22,Das2025}.
This motivates the asymptotic Laguerre--Gaussian (ALG) representation employed
below.

The fixed-target approximation is appropriate when the atom is sufficiently
localized, and its recoil is irrelevant to the measured observable. It does
not, however, resolve how the photon's linear and orbital angular momenta are
shared between the emitted electron and the residual ion. Momentum sharing in
photoionization and strong-field ionization is already known to be sensitive
to the finite nuclear mass and to the ionization regime~\cite{Chelkowski2014,
He2017}. The analogous issue for a vortex photon is more subtle because the
beam is a coherent superposition of plane-wave components with different
transverse momenta. If distinct components produce distinguishable CM recoils,
the unobserved recoil may retain which-component information and thereby alter
the coherence of the reduced electronic state. This raises the central
question of the present work: when CM motion and the finite proton mass are
included explicitly, does the optical OAM appear in the relative
electron--proton motion, in the CM recoil, or in a preparation-dependent
combination of the two? Moreover, how will the absolute motion of the electron and the proton in the lab frame be affected by the vortex light?

In this Letter, we show that the reduced electron state produced by photoionization with an optical vortex depends crucially on the preparation of the atom's initial CM state. We contrast two limiting cases: an atom with sharply defined CM momentum and an atom localized in the transverse plane. For a momentum-sharp initial CM state, each plane-wave component of the optical vortex transfers a different transverse momentum to the atom. The final CM recoil therefore carries information about which azimuthal component of the photon's momentum cone was absorbed. If the CM motion is not measured and is traced out, this recoil record suppresses the coherence between different plane-wave components of the incident vortex. Consequently, the reduced state of the electron--nucleus relative motion need not retain the vortex character, or a well-defined OAM, predicted by a fixed-target treatment. In the regime where the optical phase varies negligibly across the transverse extent of the atom, the field cannot resolve the electron and nucleus separately. The photon OAM is then transferred predominantly to the motion of the atom as a whole, i.e., to the CM degree of freedom, rather than to the relative electron--nucleus motion. This statement, however, should not be interpreted as implying that the proton carries all of the OAM essentially in the laboratory frame. The laboratory-frame OAMs of the electron and proton are not simply identified with the relative OAM (\rOAM) and CM OAM (\cmOAM), respectively. Their transformation contains an additional correlation term that can transfer a substantial fraction of the angular momentum to the electron and can even give the proton an OAM component of the opposite sign. The situation is qualitatively different for a transversely localized initial CM state. Localization requires a coherent superposition of CM momenta, including off-diagonal momentum-space coherences. These coherences prevent the recoil from providing distinguishable which-component information and thereby preserve the interference between the plane-wave components of the optical vortex. In this limit, the electron-vortex amplitude obtained in the fixed-target description is recovered. The apparent difference between the momentum-sharp and fixed-target descriptions therefore reflects the distinct treatment of the atomic center-of-mass degree of freedom rather than any physical inconsistency. They correspond to different preparations of the atomic CM state and, consequently, to different amounts of recoil information retained in the unobserved CM degree of freedom. The resulting OAM content also depends on whether one considers the relative motion, the CM motion, or the individual electron and proton motions in the laboratory frame.

\begin{figure}[b]
    \centering
    \includegraphics[width=1.0\linewidth]{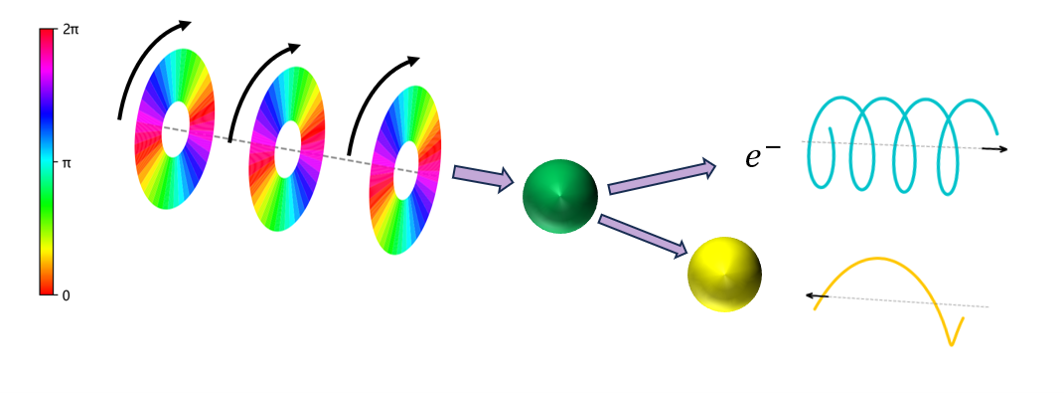}
    \caption{Schematic of the photoionization process driven by vortex light. Upon ionization of the atom (green sphere), both the photoelectron and the recoiling ion (yellow sphere) acquire OAM along the $z$-direction from the incident photon. As shown in the later discussion, the photoelectron can carry OAM exceeding the photon’s total angular momentum, while the ion acquires oppositely directed OAM, preserving the total angular momentum of the system.}
    \label{fig:concept1}
\end{figure}

We model the photoionization process as a one-photon transition from a hydrogen-like ground state to a continuum state of the electron and ion. The total Hamiltonian of the light-atom interaction is written as $H(\mathbf R,\mathbf{r},t) = H_0+H_I(\mathbf R,\mathbf
{r},t)$, where $H_0$ is the free Hamiltonian of the hydrogen-like atom with explicit CM motion, which reads $H_0
    =
    \frac{\mathbf P^2}{2M}
    +
    \frac{\mathbf p^2}{2\mu}
    -\frac{Z\mathrm{e}^2}{4\pi\epsilon_0|\mathbf r|}.$
Here, $\mathbf P$ and $\mathbf p$ are the CM and relative momenta, respectively, $M$ is the total mass of the atom, i.e.~$M=m_e+m_p$ and $\mu$ is the reduced mass of the electron and the proton, i.e.~$\mu=\frac{m_e m_p}{m_e+m_p}$. The third term is the electron--ion Coulomb potential, with ionic charge $Z\mathrm{e}$, which is independent of the CM motion. The light-atom interaction Hamiltonian is modeled in velocity gauge, i.e.~
\begin{equation}
    H_I(\mathbf R,\mathbf r, t)
    =
    -\frac{\mathrm{e}}{m_e}\mathbf A(\mathbf r_e,t)\cdot\mathbf p_e,
    \label{eq:HI}
\end{equation}
where $\mathbf A(\mathbf r_e,t)$ is the vector potential of the light field evaluated at the position of the electron $\mathbf r_e$ in the laboratory frame and $\mathbf p_e$ is the corresponding electron momentum operator \cite{Chelkowski2014}.
The direct interaction between the light field and the ion core with a positive charge is neglected because the velocity of the proton is much smaller than that of the electron.
However, the ion core is still a dynamical participant through the CM and relative coordinates.
In the interaction Hamiltonian, the vector potential $\mathbf A(\mathbf r_e,t)$ is evaluated at the electron position $\mathbf r_e=\Rvec+\beta\mathbf r$, which is the summation of the CM coordinate $\Rvec$ and the relative coordinate $\rvec$ weighted by $\beta=m_p/M$,  the proton-to-total-mass ratio. Hence, the light atom interaction Hamiltonian $H_I(\mathbf R,\mathbf r,t)$ couples the light field to both the CM and relative coordinates.
The CM motion and its effect on the light-atom interaction are the main subjects of our work, as demonstrated in Fig.~\ref{fig:concept1}.

We use a monochromatic asymptotic Laguerre--Gaussian (ALG) beam as the driving field. The vector potential can be written as,
\begin{equation}
    \mathbf A_{\rm ALG}(\mathbf x,t;\omega) =\int \frac{\dd^2\mathbf k_{\perp}}{(2\pi)^2} a_{\kappa l}(\mathbf{k}_{\perp})\mathbf A_{\kvec\sigma}(\mathbf x,t),
\end{equation}
where $\mathbf A_{\mathbf k,\lambda}(\mathbf x) = \hat{\epsilon}_{\mathbf k \lambda}\mathrm{e}^{\mathrm{i}\mathbf{k}\cdot \mathbf{x}-\mathrm{i}\omega t}$ is the unit vector potential of a plane wave with wavevector $\mathbf k$ and helicity $\lambda$. Here, $\kvec=\kvec_{\perp}+k_z\hat{z}$, with $\kvec_{\perp}$ and $k_z\hat{z}$ denoting the transverse and longitudinal components, respectively. The transverse spectral function  
    $a_{\kappa\ell}(\mathbf k_\perp)
=
\mathcal A_{\kappa\ell}\delta(k_\perp-\kappa)\mathrm{e}^{\ii\ell\Phi}$,
where $\Phi$ is the azimuthal angle of $\mathbf k_\perp$ in the lab frame. The normalization factor $\mathcal A_{\kappa\ell}$ is a constant proportional to the total photon flux of the beam and is irrelevant in the subsequent analysis. The ALG beam carries a total angular momentum of $\ell\hbar$ per photon. We define the cone angle $\theta_k$ through $\tan\theta_k=\kappa/k_z$, which characterizes the angular spread, or divergence, of the constituent plane-wave components about the propagation axis. Small $\theta_k$ corresponds to a weakly divergent, nearly paraxial beam, whereas large $\theta_k$ describes a strongly nonparaxial beam with substantial transverse photon momentum. The complete details of the ALG beam are given in the Supplementary Material (SM).

The photoionization process is described by the transition amplitude,
\begin{equation}
    S_{fi}^{(1)}
    =
    -\frac{\ii}{\hbar}
    \int_{-\infty}^{\infty}\dd t\,
    \bra{f}\tilde H_I(\mathbf R,\mathbf r,t)\ket{i},
    \label{eq:S_matrix}
\end{equation}
where $\tilde H_I(\mathbf R,\mathbf r,t)$ is the $H_{I}(\mathbf R,\mathbf r,t)$ in the interaction picture, and $\ket{i}$ ($\ket{f}$) is the initial (final) state of the system. We generally express the initial and final states as the direct product of the CM and relative states.
Likewise, for $\ket{f}$, we assume that the relative motion is the direct product of free relative motion and the free CM motion. We denote it as $\ket{\mathbf P, \mathbf p}$. In our model, we neglect the Coulomb interaction between the outgoing electron and the ion core in the final state, i.e., we work under the strong-field approximation (SFA).

We start our calculation with an initial state consisting of the momentum eigenstates of the CM motion and the electronic ground state, namely $\ket{\mathbf P_{i}, 1s}$. Other initial CM states can be expressed as a superposition or a mixture of the momentum eigenstates. Also, because of the large mass of the ionic core, changes in the CM kinetic energy are negligible compared with the electron excess energy.

After solving the radial momentum integral, the outgoing state reads
\begin{equation}
\begin{aligned}
    \ket{\Psi_{\kappa\ell}^{(1)}(\Pvec_i)}
    =
    &{\cal N}_{\kappa\ell}
    \int_0^{2\pi}\dd\Phi\,
    \int_{\vert \pvec\vert=p_0}\dd\Omega_{\hat{\pvec}}\,\mathrm{e}^{\ii\ell\Phi}
    {\cal F}_\lambda(\hat{\pvec},\Phi;\mathbf P_i)\\&\times
    \ket{\Pvec_i+\hbar\kvec(\Phi)}_{\rm CM}
    \ket{p_0\hat{\pvec}}_{\rm rel},
\end{aligned}
    \label{eq:finalPi}
\end{equation}
where ${\cal F}_\lambda(\hat{\pvec},\Phi;\mathbf P_i) = \boldsymbol\epsilon_{\mathbf k\lambda}\cdot
(\mathbf p+\alpha\mathbf P_i)
\widetilde\phi_{1s}(\mathbf p-\beta\hbar\mathbf k(\Phi))$ is an internal factor. In Eq.~\eqref{eq:finalPi}, $p_0 = \sqrt{2m_e(\hbar\omega-I_p)}$ and $\kvec(\Phi) = (\kappa \cos{\Phi},\kappa\sin{\Phi},k_z)$. The coefficient $\alpha = m_e/M$ is the electron-to-total-mass ratio (see SM).

\begin{figure}
    \centering
    \includegraphics[width=0.7\linewidth]{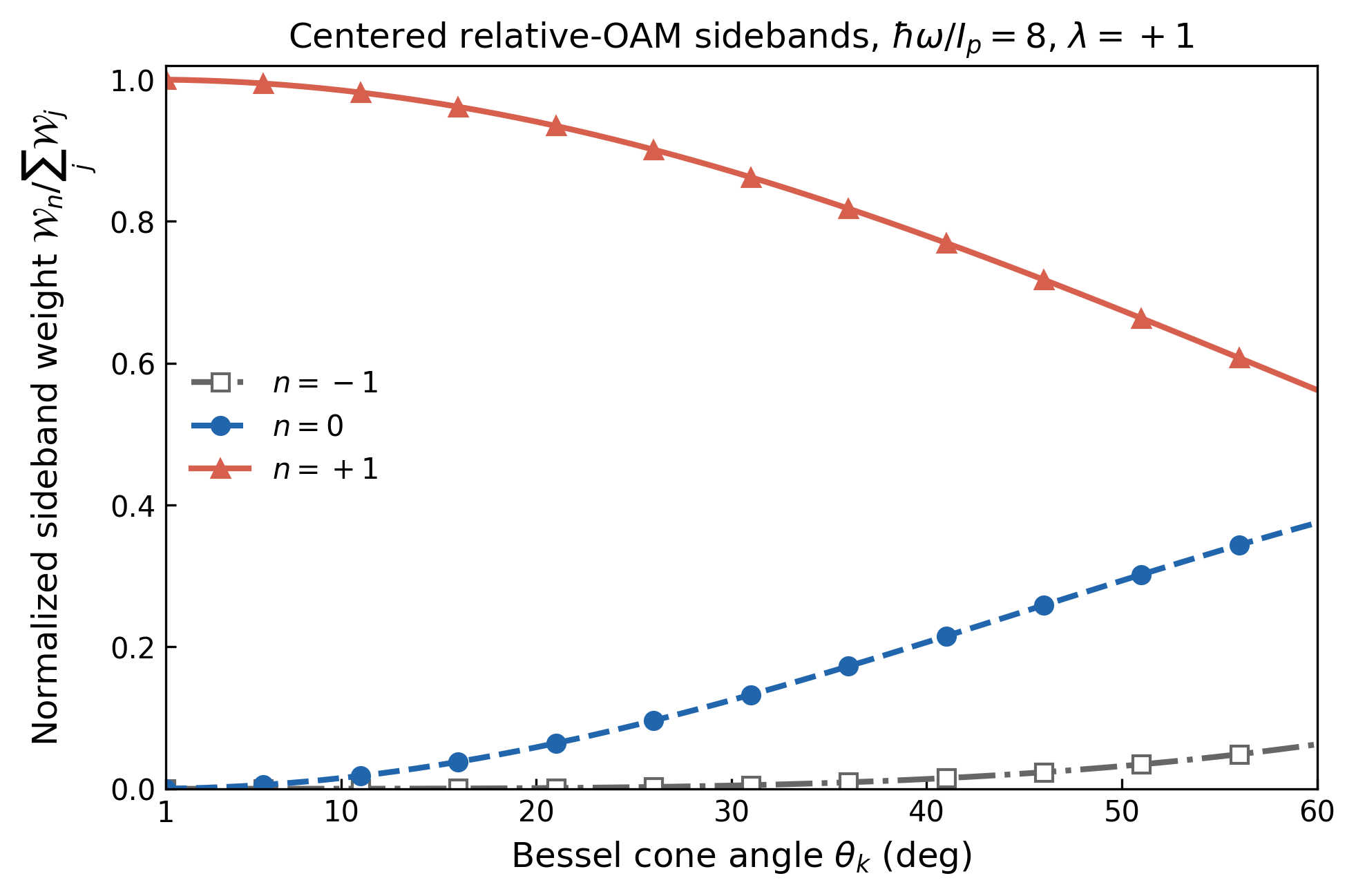}
    \caption{Leading sideband weights as a function of $\theta_k$. As $\theta_k$ increases, the weight of the dominant $n=1$ channel is progressively redistributed to the $n=0$ and $n=-1$ channels. The photon helicity is fixed at $\lambda=+1$.}
    \label{fig:leading_sideband_weights}
\end{figure}

To solve the \rOAM, we expand ${\cal F}_\lambda(\hat{\pvec},\Phi;\mathbf P_i)$ in terms of the eigenfunctions of $L_{{\rm rel},z}=-\ii\hbar\,\partial/\partial\phi_p$, where $\theta_p$ and $\phi_p$ are the relative polar and azimuthal angles of the photoelectron momentum, respectively:
\begin{equation}
\mathcal F_\lambda (\hat{\mathbf p},\Phi;\mathbf{P}_i)
=\sum_{n=-\infty}^{\infty}
\sum_{j=-1}^{1}
\Gamma_{n,j}^{(\lambda)}(\theta_p)
\ee^{\ii n\phi_p}\ee^{-\ii(n+j)\Phi}.
\label{eq:FnExpansion}
\end{equation}
In Eq.~\eqref{eq:FnExpansion}, the indices $j=0$ and $j=\pm1$ imply that the amplitude $\Gamma_{n,j}^{(\lambda)}$ is a function of $\alpha P_{i,z}$, and $\alpha P_{i,\perp}$, respectively. Only $\Gamma_{n,0}^{(\lambda)}$ does not vanish when $\alpha\rightarrow0$. The explicit form and the derivation of $\Gamma_{n,j}^{(\lambda)}(\theta_p)$ are given in the SM. Next we define the sideband weight of the $n$-th OAM channel as
\begin{equation}
\mathcal W_n^{(\lambda)}(\mathbf P_i)
=\int_0^\pi \dd\theta_p\,\sin\theta_p
\sum_{j=-1}^{1}
\left|
\Gamma_{n,j}^{(\lambda)}(\theta_p;\mathbf P_i)
\right|^2.
    \label{eq:Wn}
\end{equation}
The sideband weight $\mathcal W_n^{(\lambda)}(\mathbf P_i)$ gives the probability for the outgoing electron to occupy the $n$-th OAM sideband. This representation follows naturally from the decomposition of the photoelectron state into a discrete set of OAM sidebands. Since the initial relative momentum vanishes for the ground state, the normalized expectation value of the electron OAM is obtained as the weighted average over all sidebands, i.e.
\begin{equation}
    {\langle L_{{\rm rel},z}\rangle}
=\hbar{\displaystyle\sum_{n=-\infty}^{\infty}
n\mathcal W^{(\lambda)}_n(\mathbf P_i)}
/{\displaystyle\sum_{n=-\infty}^{\infty}
\mathcal W^{(\lambda)}_n(\mathbf P_i)}.
    \label{eq:Lrel}
\end{equation}
In the low-photon-momentum limit, the dominant contributions arise from the three leading sidebands, $W_{-1}$, $W_0$, and $W_{1}$ (see SM). For the results shown in Fig.~\ref{fig:leading_sideband_weights}, the photon energy is fixed at $\hbar\omega=8I_p$, with $I_p=0.5$ a.u. for hydrogen. We consider hydrogen atoms confined in a large chamber and described by a thermal-equilibrium distribution, corresponding to an incoherent mixture of CM momentum eigenstates. 

Next, we consider the absolute motion in the laboratory frame. In photoelectron measurements, the electron momentum is observed in the lab frame rather than in the relative frame.
In addition to resolving the \rOAM\; and the \cmOAM, the OAM of the photon-electron and the ion core in the lab frame must be considered. The electron's OAM  $\hat{L}_{e,z}$ and the ion core's OAM $\hat{L}_{{p},z}$ operators read
\begin{equation}
    \begin{aligned}
        \hat{L}_{e,z}
&=\alpha \hat{L}_{{\rm CM},z}+\beta \hat{L}_{{\rm rel},z}
+\widehat{\mathcal C}_z\\
\hat{L}_{p,z}
&=\beta \hat{L}_{{\rm CM},z}+\alpha \hat{L}_{{\rm rel},z}
-\widehat{\mathcal C}_z.
    \end{aligned}
    \label{eq:ep_angular_momentum}
\end{equation}
where $
    \widehat{\mathcal C}_z
=(\hat{\mathbf {R}}\times\hat{\mathbf p})_z
+\alpha\beta(\hat{\mathbf r}\times\hat{\mathbf P})_z$
is the correlated angular momentum between the relative motion and the CM motion, though it does not follow the commutator of angular momentum operators. The correlated angular momentum is the sum of two parts, namely $\widehat{\mathcal C}_z^{(R p)}
=(\mathbf R\times\mathbf p)_z$ and $\widehat{\mathcal C}_z^{(r P)}
=\alpha\beta(\mathbf r\times\mathbf P)_z$. 
We also apply the sideband structure to $\widehat{\mathcal C}_z^{(R p)}$ and $\widehat{\mathcal C}_z^{(r P)}$ and use series to solve their expectation value.  This formulation is not only computationally convenient but also provides a channel-resolved decomposition of the correlated angular momentum, making it possible to identify the contribution of each relative-motion channel. It is worth noting that, in the heavy-core approximation where $\alpha\rightarrow 0$, $\widehat{\mathcal C}_z^{(R p)}$ does not vanish in general (see SM).

A momentum eigenstate has infinite spatial uncertainty and therefore no well-defined spatial center. Consequently, the expectation values of the absolute electron and proton OAM, $\langle \hat{L}_{e,z}\rangle$ and $\langle \hat{L}_{p,z}\rangle$, are not well defined for such a state. To investigate how the angular-momentum partition \AMp\, depends on the spatial localization of the atom, we therefore replace the initial CM momentum eigenstate $\ket{\mathbf P_i}$ by a Gaussian CM wavepacket of finite spatial width. Relative to the momentum-eigenstate case, the finite localization of the CM introduces a modulation of the OAM-sideband weights. In the heavy-core limit, $\alpha\rightarrow0$ and $\beta\rightarrow1$, the resulting sideband weights take the form $\mathcal W_n^{G,\infty}=8\pi^3 W_n^{(\lambda)}E_{\ell-n}(a)$, where $a=\kappa\sigma_R$ and $E_m(a)=e^{-a^2}I_{|m|}(a^2)$, with $I_m(z)$ denoting the modified Bessel function of the first kind. Thus, the localization of the CM modifies the relative importance of the different OAM sidebands through the dimensionless parameter $\kappa\sigma_R$.

The expectation value of the correlated angular momentum originates from the coherence between neighboring relative-OAM channels, in particular between the $n$th and $(n+1)$th sidebands. It therefore remains, in general, nonzero even in the heavy-core limit (see SM).

\begin{figure*}[t!]
    \centering
    \subfloat[]{\includegraphics[width=0.3\linewidth]{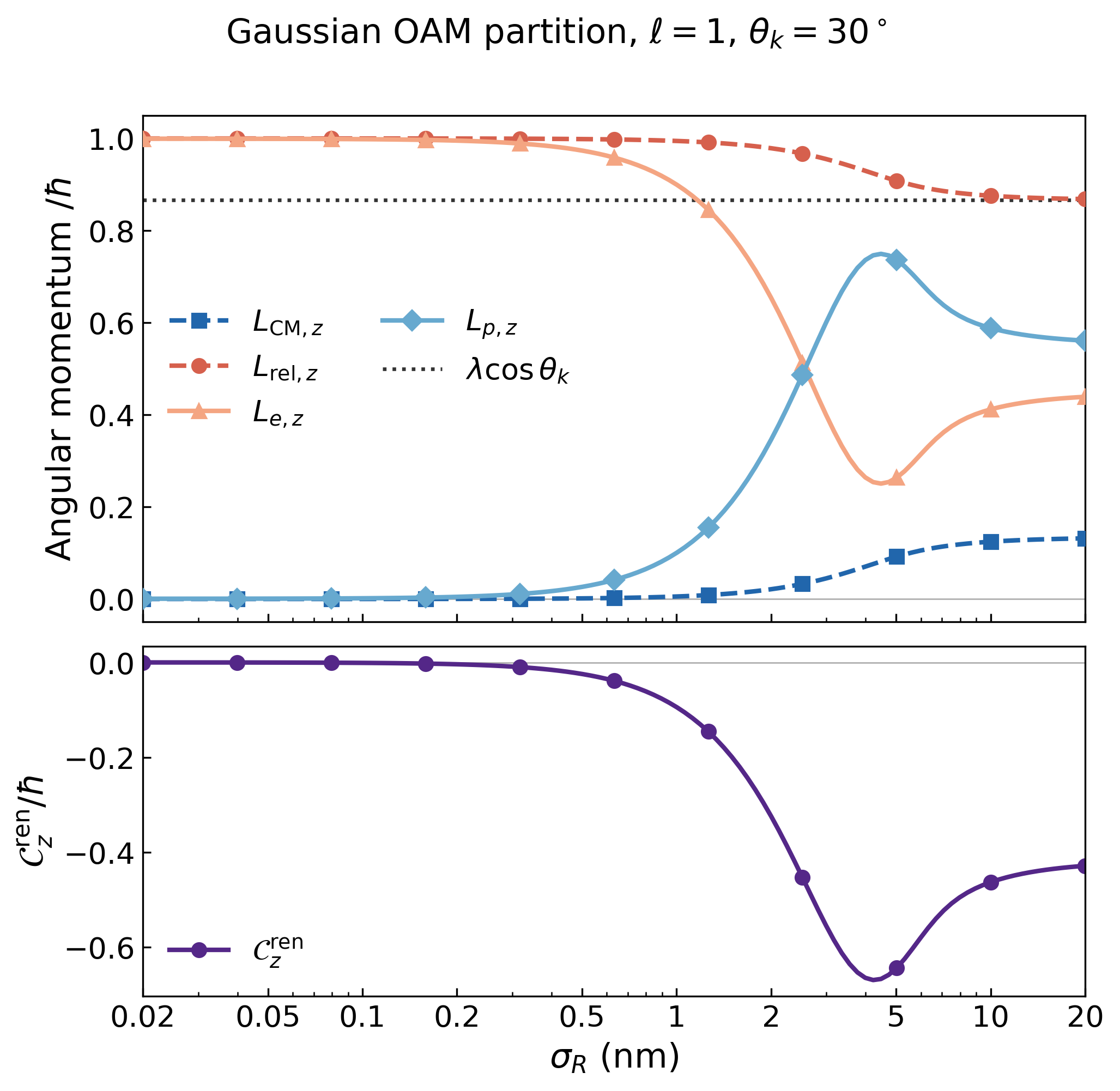}}\subfloat[]{\includegraphics[width=0.3\linewidth]{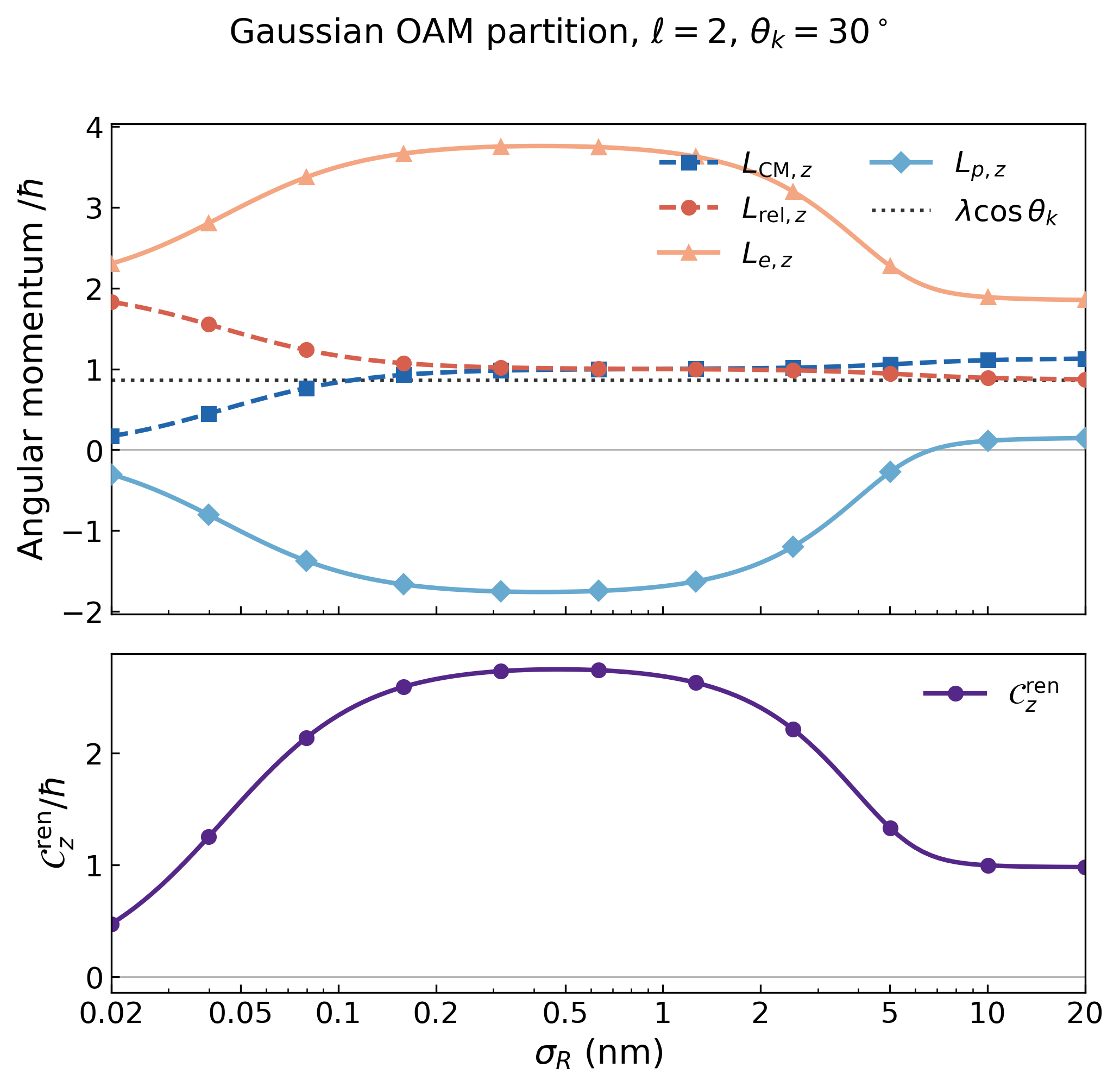}}\subfloat[]{\includegraphics[width=0.3\linewidth]{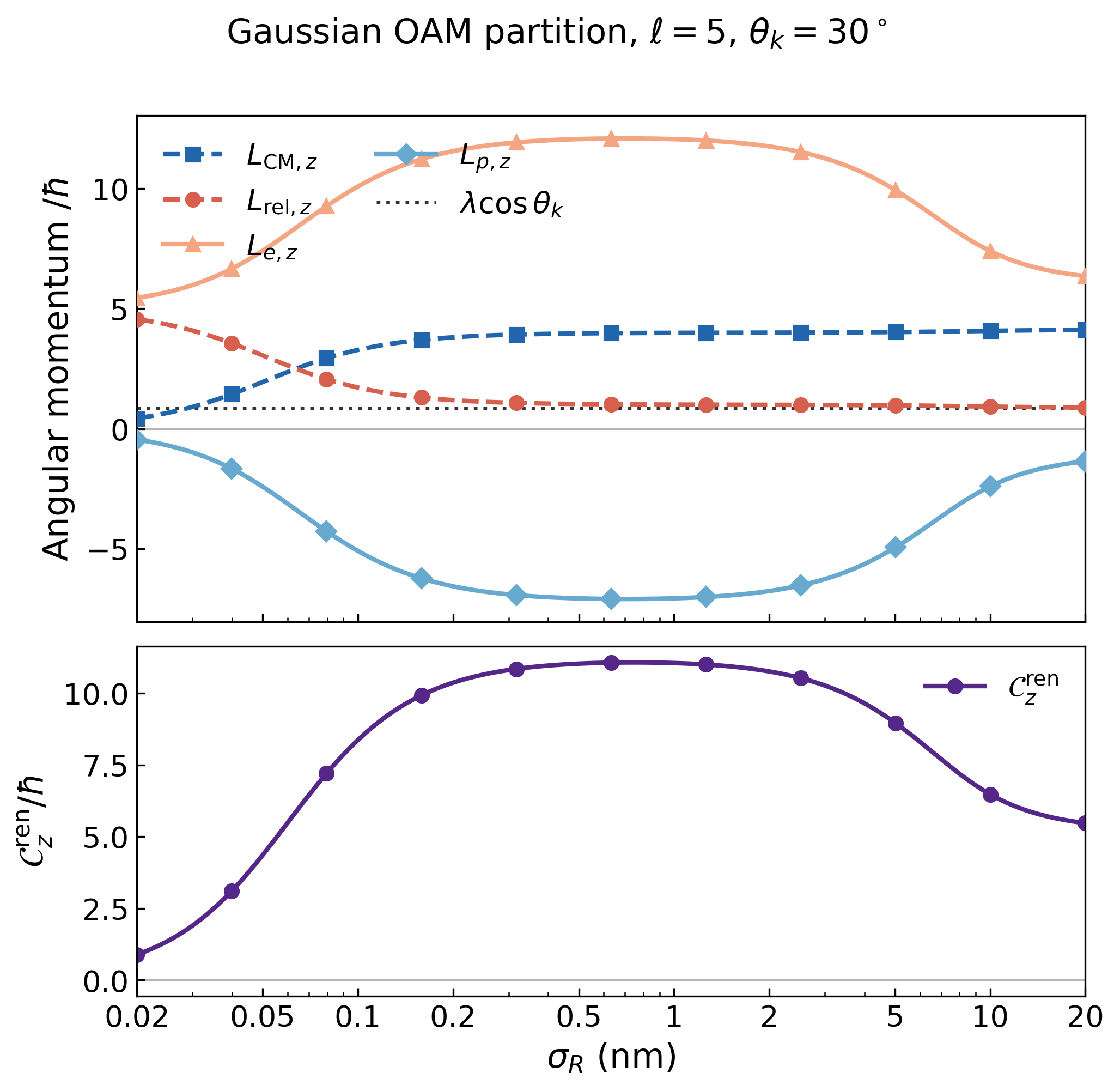}}
    \caption{The average value of the \rOAM, the \cmOAM, the \eOAM, the proton's OAM, and the correlated angular momentum operator as functions of the spatial uncertainty $\sigma_R$. The total angular momentum indices are chosen as $\ell =1,2,5$ and the cone angle is chosen as $\theta_k =30^\circ$. Other settings are the same as Fig.~\ref{fig:leading_sideband_weights}.}
    \label{fig:Gaussian_OAM_partition}
\end{figure*}

We numerically consider a CM wavepacket with vanishing mean momentum and a spatial uncertainty in the nanometer range. A representative example of the resulting OAM partition is shown in Fig.~\ref{fig:Gaussian_OAM_partition}. In the strong-localization limit, $\sigma_R\rightarrow0$, one finds $\langle \hat{L}_{e,z}\rangle=\langle \hat{L}_{\mathrm{rel},z}\rangle\rightarrow\ell\hbar$, recovering the fixed-target result of Ref.~\cite{Das2025}. This limit, however, corresponds to an extremely strong confinement of the atomic CM and is therefore distinct from the spatially delocalized momentum-eigenstate preparation.

We have investigated how the optical OAM is partitioned in the photoionization of a hydrogen-like atom by vortex light, considering targets prepared in a CM momentum eigenstate, in thermal equilibrium, and as spatially localized wavepackets. Our results show that the final OAM partition is not determined by the optical mode alone, but depends crucially on the initial quantum state of the atomic center of mass. For an initial CM momentum eigenstate, the azimuthal component of the absorbed photon momentum is encoded in the atomic recoil. Tracing over this unobserved recoil removes the coherence between different plane-wave components of the vortex field that is required to form a pure electron-vortex state. In the scalar, small-retardation limit, the optical OAM is consequently transferred predominantly to the CM degree of freedom. By contrast, for a spatially localized target, the initial state contains a coherent superposition of CM momenta. The coherent sum over these momentum components generates the phase $\mathrm{e}^{\ii\kvec(\Phi)\cdot\Bvec}$ and progressively restores the fixed-target vortex-electron amplitude. For the more general case of an initial Gaussian CM wavepacket, the OAM partition is controlled primarily by the spatial width of the atomic wavepacket and by the cone angle of the vortex beam. Electron-vortex formation therefore requires not only transverse structuring of the optical field, but also sufficient localization of the atomic target on the relevant transverse length scale. In the limiting case of a CM wavefunction approaching a position eigenstate at the vortex center, together with the heavy-core limit $M\rightarrow\infty$, the electron and relative coordinates become effectively equivalent and the outgoing electron approaches the $\ell\hbar$ OAM channel obtained in fixed-target calculations \cite{Das2025}. In this limit, the CM dynamics becomes negligible and the internal motion coherently samples the full azimuthal structure of the vortex field. The fixed-target Bessel-electron picture and the CM-resolved recoil picture should therefore be understood as complementary limits of the same photoionization process. More generally, the observation and interpretation of OAM transfer in vortex-light photoionization require control over or characterization of the initial target localization and the coherence retained in the recoil degree of freedom, rather than consideration of the outgoing-electron distribution alone.

\section*{Acknowledgments}
This work is supported by the National Key Research and Development Program of China (Grant No. 2023YFA1406801), and the National Natural Science Foundation of China (Grant No. 12595343, 12674326). M.~F.~C.~acknowledges support by the Quantum Science and Technology-National Science and Technology Major Project (Grant No. 2025ZD0301000),  the National Key Research and Development Program of China (Grant No. 2023YFA1407100), the Guangdong Province Science and Technology Major Project (Future functional materials under extreme conditions - 2021B0301030005), the National Natural Science Foundation of China (Grant No. 12574092) and from the Knut and Alice Wallenberg Foundation through the Wallenberg Centre for Quantum Technology (WACQT).

\bibliographystyle{apsrev4-2}
\bibliography{references}

\end{document}